\documentclass{vgtc}                          

\ifpdf
  \pdfoutput=1\relax                   
  \usepackage{graphicx}                
  \DeclareGraphicsExtensions{.pdf,.png,.jpg,.jpeg} 
\else
  \usepackage{graphicx}                
  \DeclareGraphicsExtensions{.eps}     
\fi%

\graphicspath{{figures/}{pictures/}{images/}{./}} 
\usepackage{mathtools}
\usepackage{enumitem}
\usepackage{microtype}                 
\PassOptionsToPackage{warn}{textcomp}  
\usepackage{textcomp}                  
\usepackage{mathptmx}                  
\usepackage{times}                     
\usepackage{cite}                      
\usepackage{tabu}                      
\usepackage{booktabs}                  

\onlineid{0}

\vgtccategory{Research}

\vgtcinsertpkg

\title{Let's Gamble: How a Poor Visualization Can Elicit Risky Behavior}

\author{Melanie Bancilhon\thanks{e-mail: mbancilhon@wustl.edu}\\ %
        \scriptsize Washington University in St. Louis %
\and Zhengliang Liu\thanks{zhengliang@wustl.edu}\\ %
        \scriptsize Washington University in St. Louis %
\and Alvitta Ottley\thanks{alvittao@wustl.edu}\\ %
     \scriptsize Washington University in St. Louis }

\teaser{
  \centering
  \includegraphics[width=.6\linewidth]{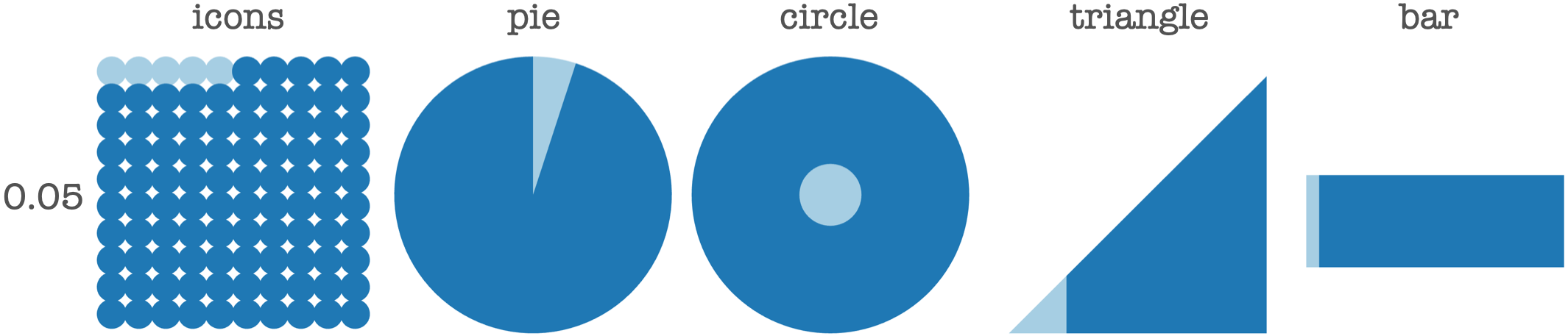}
  \caption{The visualization conditions used in the study. Although we had no control over the display size, the default size of the visualizations were $200\times200$ pixels, and the colors were colorblind safe.}
  \label{fig:conditions}
}

\abstract{Data visualizations are standard tools for assessing and communicating risks. However, it is not always clear which designs are optimal or how encoding choices might influence risk perception and decision-making. In this paper, we report the findings of a large-scale gambling game that immersed participants in an environment where their actions impacted their bonuses. Participants chose to either enter a lottery or receive guaranteed monetary gains based on five common visualization designs. By measuring risk perception and observing decision-making, we showed that icon arrays tended to elicit economically sound behavior. We also found that people were more likely to gamble when presented area proportioned triangle and circle designs. Using our results, we model risk perception and discuss how our findings can improve visualization selection.%
} 

\CCScatlist{
  \CCScatTwelve{Human-centered computing}{Visu\-al\-iza\-tion}{};
  \CCScatTwelve{Decision Theory}{Risk Behavior}{Evaluation methods}{}
}

\begin{document}


\firstsection{Introduction}
\maketitle
\label{sec:introduction}


There are several competing views on what it means to make a decision. Such choices include deciding whether to bike or drive based on the chance of rain, whether to opt for preventive health care based on the likelihood of developing a disease or whether to enter a gamble based on the chance of winning a prize. 
Psychologists believe that we make choices based on empirical evidence and beliefs about the likelihood of specific events~\cite{tversky1992}.
Economic theorists view decision-making as a selection between alternatives based on a weighted sum of probabilities~\cite{prelec1991}. 
In many cases, it is increasingly common to use data visualization to support reasoning about risks and to aid sound decision-making~\cite{padilla2018decision}.

There is a wide sampling of work that investigates a variety of visualization designs for communicating risks. Researchers have used real-life scenarios such as catching a bus, playing a battleship game or predicting a weather forecast to evaluate how visualizations such as dotplots and other elementary charts such as line graphs affect decisions \cite{correll2018,greis2018,kay2016ish}. 
At the same time, we have seen an increased prevalence of visualization in communicating statistical information to the general public. For example, the `flatten the curve' visualization helped shape the public lexicon during the COVID-19 pandemic. This phenomenon demonstrates that visualization can have a significant impact on society, especially when decisions are life-altering. However, a designer can represent the same data using different, yet equally theoretically valid visualization designs~\cite{mackinlay2007show}, and it is sometimes difficult for designers to identify the best fit for the data~\cite{pandey2015deceptive}. 

The goal of this paper is to establish clear guidelines for visualization selection by investigating how design choices can impact decision-making under risk.
We utilize a real-life gambling game with financial incentive to investigate the effect of five visualization designs on decision-making under risk. Consider the following hypothetical gamble~\cite{kahneman2013prospect}:

\begin{quote}\label{example-prospect}
\begin{flushleft}
\textit{Which do you prefer?}\newline
\textbf{A:} 50\% chance to win \$1000, 50\% chance to win \$0\newline
\textbf{B:} \$450 for sure\newline
\end{flushleft}
\end{quote}

\vspace{-7.15mm}
\noindent 
\newline
Decision theorists have long studied simple gambles such as this because they provide a straightforward framework that shares key characteristics with complex real-world situations~\cite{bruhin2010risk}. 
The lottery scenario is a useful test-bed because one need only to weigh the risk of entering the lottery against the possible return associated with the guaranteed payment. 
By modeling decision-making behavior for each visualization design, our work provides a basis for visualization selection. We make the following contributions:

\begin{itemize}
    \item We demonstrate the effect of probability distortion on risk behavior. Participants in the visualization groups were more risk seeking with small probabilities and more risk averse with large probabilities.
    \item  We show that visualization design influences risk behavior. \textit{Icon} lead to the most economically rational decision-making while \textit{triangle} and \textit{circle} supported more risk-averse decisions.

\end{itemize}

\section{Background}
Risk taking has been studied by scholars within a broad range of fields, from business to engineering to health care and developmental education. The consequences of people's actions in the presence of risk are often dramatic, hence the importance of investigating how risk is communicated and perceived ~\cite{yates1992}. In everyday domains that involve uncertainty, visualization is often used as a tool to communicate risks.

In the medical field, for example, Galesic et al. conducted an experiment where patients were shown the risk of a disease in either a numerical or a visual format~\cite{galesic2009}. Participants in the study were asked to rate the seriousness of the disease and the importance of screening on a scale of 1 to 15. The results showed that participants who saw the information in a numerical format rated the disease as more serious than it really is compared to the group who were shown icon arrays and considered screening more important. 
They stated that visual aids help patients to make more informed medical decisions~\cite{galesic2009}. Similarly, Ruiz et al. conducted a study where they asked at-risk patients to decide whether they would opt for screening based on hypothetical risk information about the disease ~\cite{ruiz2013}. They confirmed the prior finding that people are more risk-averse when presented with icon arrays. 


There is extensive literature on risk communication and decision-making in the visualization community. Kay et al. used visualization to communicate uncertainty of transit data~\cite{kay2016ish}. They found that quantile dotplots lead to significantly better outcomes when making decisions about when to catch a bus. 
While surveys and hypothetical scenarios have often been used in the medical field, some researchers such as Kay have observed users in action by immersing them in real-life simulations. Bisantz et al. used a missile defense game to evaluate how well users make decisions under uncertainty~\cite{bisantz2011}. Moreover, Fernandes et al. who built on Kay's work argued that decisions are best observed when using financial incentive~\cite{fernandes2018}. 
In their study, participants were compensated based on how well they minimized their wait time at the bus stop while still being on time to catch their bus. 

In this paper, we adopted a similar approach where participants engage in realistic decision-making and are compensated according to their choices. 
In particular, we situate our approach to investigate how visualization influences decision-making in \textit{Decision Theory}. We created a real-life gambling scenario and observed participants' lottery decisions to provide a framework that can improve visualization selection to assist decision-making in a number of areas.

 
\subsection{Decision Theory}
Economists and psychologists have long studied how people make choices under risk by investigating prospects or gambling scenarios. A prospect is a contract:
\begin{equation}
    (x_1, p_1:...:x_n, p_n),
\end{equation}
\noindent
that yields $x_i$ with probability $p_i$, where $\sum_{i=1}^{n} p_i = 1$~\cite{kahneman2013prospect}. Prospects provide a simple model for understanding risky decisions. 

The classical method for evaluating a gamble is through assessing its expected value. The expected value of a prospect is the sum of the outcomes where the probabilities weight each value:
\begin{equation}
    ev = \sum_{i=1}^{n} p_ix_i
\end{equation}
\noindent
For example, consider the gambling scenario in section 2.1, the expected value of option A is 500 ($ev = .5 \times 1000 + .5 \times 0 $) and the expected value of option B is 450 ($ev = 1 \times 450$). A \textit{rational} decision-maker would then choose option A over option B. However, most people would choose the sure payment of \$450. This highlights the perhaps obvious conjecture that humans are not always rational.

One of the dominant theories of decision making, \textit{Expected Utility Theory} (EUT), has served for many years as both a model that describes economic behaviors~\cite{friedman1948utility} and a model of rational choice~\cite{keeney1993decisions}.  In particular, it states that people make choices based on their \textit{utility} - the psychological values of the outcomes. 
Using EUT, we can assess the overall utility of a gamble by summing the utilities of the outcomes weighted by their probabilities. 
\begin{equation*}
    U = \sum_{i=1}^{n} p_i u(x_i)
\end{equation*}
\noindent
This model, however, still assumes that most humans are rational and consistent, and solely decide on prospects based on their utility~\cite{kahneman2013prospect}.

A refinement of EUT that is used to describe risk perception and decision-making with risk empirically is known as \textit{Prospect Theory}~\cite{kahneman2013prospect}. The theory posits that people tend to underweight common or high-frequency events while over-weighting rare or low-frequency events. Typically, there is a probability weighting function $\pi$ where $\pi$(p)\textgreater p  \text{ when p is small and}  $\pi$(p)\textless p,  \text{ when p is large but not a certainty}.

\noindent
$\pi$ reflects the subjective desirability of a choice, which in practice replaces the stated probabilities with weighting factors $\pi(p)$. 
Furthermore, Prospect Theory stipulates that such a phenomenon has a two-fold impact on binary decision-making: (1) people tend to favor the option of getting a large gain with a small probability over getting a small gain with certainty, and (2) people tend to prefer a small loss with certainty over a large loss with tiny probability. 
Lotteries can be expressed in terms of gains or losses.  For our purpose, we limit our scope to the gain domain. 

Relevant to the current work, Bruhin et al.~\cite{bruhin2010risk} conducted a series of large scale lottery studies and classified the distributions of behavioral types of different portions of the population based on how closely their behaviors are to that described by EUT and Prospect Theory. 
They analyzed the Relative Risk Premia, a descriptive metric of how risk-seeking or risk-averse a choice is~\cite{bruhin2010risk}. They showed that participants were risk-seeking for low-probability gains and risk-averse for high-probability gains. 

\section{Research Question}
This paper extends the prior research by investigating the complex relationship between visualization design, risk perception, and decision-making.
Prior work in psychology and economic theory provides a convenient framework to investigate this relationship. In particular, we leverage a classical task for eliciting decision-making under risk by observing actions as participants chose between entering a gamble and receiving a guaranteed bonus payoff. We used five common designs and a text condition to display seven lottery probability values that ranged from a 5\% to a 95\% chance of winning. We framed the following research question to guide our investigation:

\begin{quote}
    \textit{Does visualization impact decision-making under risk?}
\end{quote}

    

\begin{table}[t!]
\caption{The prospects that were used in the study. $p_1$ denotes the probabilities ($p_2 = 1 - p_1$) and $x_1$ and $x_2$ are the outcomes. Participants saw all combinations in a random order.}
\label{tab:prospects}
\resizebox{.95\textwidth}{!}{\begin{minipage}{\textwidth}
\begin{tabular}{|llll|lllll|llll|}
\hline
$p_1$   & $x_1$  & $x_2$ &  &  & $p_1$   & $x_1$  & $x_2$ &  &  & $p_1$   & $x_1$ & $x_2$ \\ 
\hline
.05 & 20  & 0  &  &  & .25 & 50  & 20 &  &  & .75 & 50 & 20 \\
.05 & 40  & 10 &  &  & .50  & 10  & 0  &  &  & .90  & 10 & 0  \\
.05 & 50  & 20 &  &  & .50  & 20  & 10 &  &  & .90  & 20 & 10 \\
.05 & 150 & 50 &  &  & .50  & 40  & 10 &  &  & .90  & 50 & 0  \\
.10  & 10  & 0  &  &  & .50  & 50  & 0  &  &  & .95 & 20 & 0  \\
.10  & 20  & 10 &  &  & .50  & 50  & 20 &  &  & .95 & 40 & 10 \\
.10  & 50  & 0  &  &  & .50  & 150 & 0  &  &  & .95 & 50 & 20 \\
.25 & 20  & 10 &  &  & .75 & 20  & 0  &  &  &     &    &    \\
.25 & 40  & 10 &  &  & .75 & 40  & 10 &  &  &     &    &   \\
\hline
\end{tabular}
\end{minipage}}
\end{table}

\section{Experiment}
\begin{figure}[t]
    \centering
    \includegraphics[width=.4\textwidth]{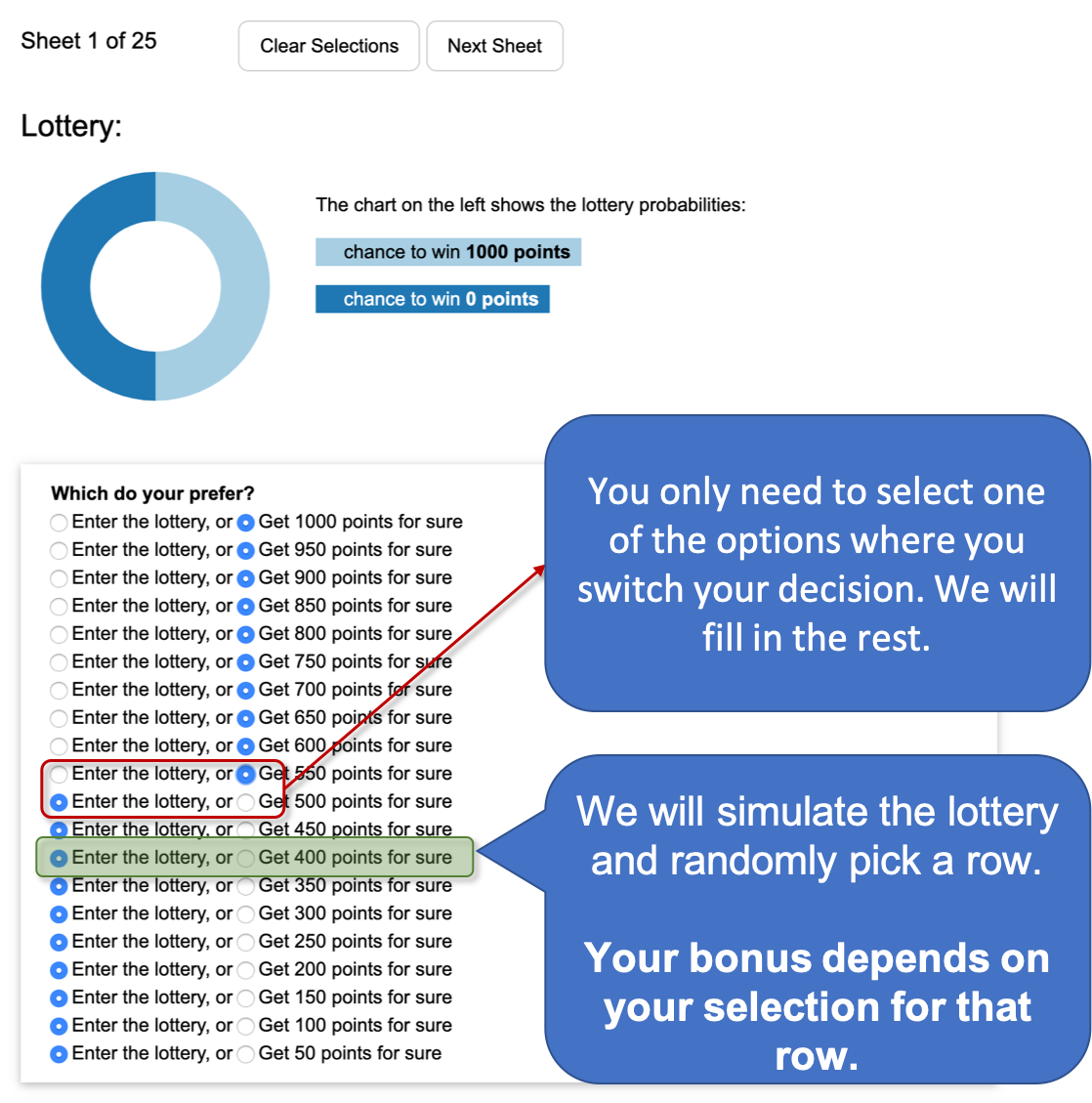}
    \caption{The example lottery sheet used in the study, with a different visualization and reward prize. Participants only needed to select the row where they switched either decision, as the system automatically populated the remaining radio buttons.}
    \label{fig:lotterysheet}
\end{figure}

Replicating the experiment design of prior work in the economic decision-making domain~\cite{bruhin2010risk}, we presented participants with two-outcome lotteries that were choices between risky and certain gains. We used a points system for our payoff quantities where 1 point equaled \$0.01. The probabilities, $p_i$, were drawn from the set $P = \{.05, .1, .25, .5, .75, .9, .95 \}$ and the outcomes $x_1$ and $x_2$ ranged from 0 to 150 points (\$0 to \$1.50). Table~\ref{tab:prospects} summarizes the probability and outcome combinations used in the study. Each lottery sheet comprised of a list of 20 equally-spaced outcomes that ranged from $x_1$ to $x_2$. Figure~\ref{fig:lotterysheet} shows an example of the lottery sheets.

\subsection{Visualization Designs}
We selected visualization designs that employed different encodings to represent data~\cite{cleveland1984graphical}. In particular, \textit{icon arrays} are used frequently in the medical community ~\cite{galesic2009,micallef2012assessing,ottley2015improving}. 
\textit{Pie} and circle charts are common in information graphics and market surveys~\cite{skau2016arcs,kosara2019impact}.
The \textit{triangle}, \textit{circle}, and \textit{bar} are all used in the medical community~\cite{vizhealth}, and represent area and length judgments~\cite{cleveland1984graphical}. 

\subsection{Participants}

We recruited 300 participants from Amazon Mechanical Turk. There were 193 men and and self-reported ages ranged from 19 to 65 years ($\mu= 34.0;\sigma=9.01$). $56\%$ of our participants self-reported to have completed at least a college education. Each participant had a HIT approval rate of 98\% with at least 100 approved HITs. We did not limit geographical location, but we required all subjects to be English-speaking between 18 and 65 years. We paid a base rate of \$1.00, plus a bonus of up to \$10.70 depending on the lottery outcomes. Bonus payments range from \$4.35 to \$9.30 ($\mu = \$7.22 ; \sigma = \$0.94$). Participants worked at their own speed and the average completion time was approximately 29 minutes. 

\subsection{Procedure}
After selecting the task on Mechanical Turk, participants consented per Washington University's IRB protocol. 
Then they saw a short tutorial that included one trial text-only round and explained the selections and the bonus calculation using the image shown in Figure~\ref{fig:lotterysheet}. To prevent potential biasing, we used a donut chart for the instructions which was not a visualization condition in the study.
Each participant was randomly assigned one of the six visualization conditions. They were presented with 25 two-outcome lotteries that were choices between risky and certain gains. There were no default selections and the order of the sheets was counterbalanced to prevent ordering effects.

\subsection{Measures}
The experiment included the following independent variables:
\begin{itemize}
    \item \textbf{7 Probability Values:} $\{.05, .1, .25, .5, .75, .9, .95 \}$
    \item \textbf{6 Visualizations:} $\{icon, pie, circle, triangle, bar, none\}$
\end{itemize}

\noindent
To measure decision quality, our dependent variable was:

\begin{itemize}

    \item \textbf{RRP: } The Relative Risk Premia is used to evaluate the quality of the lottery decisions~\cite{bruhin2010risk} and can be seen as a measure of rationality. 
    \[
        RRP = (ev - ce )/ |ev|, 
    \]
    where ev denotes the expected value of the lottery outcome and ce is the certainty equivalent of the lottery. We can calculate the utility of a given prospect using equation (2) from section 2.2. We calculate the lottery's certainty equivalent as the average of the smallest certain amount that the participant selected on the sheet and the subsequent certain amount on the sheet. For example, let us assume that figure~\ref{fig:lotterysheet} indicates a participant's selection. The certainty equivalent here is 525 ( $ce = (550 + 500) / 2$ ). 
    $RRP > 0$ indicates risk aversion, $RRP < 0$ implies risk seeking behavior and $RRP = 0$ suggests risk neutrality.
    
\end{itemize}

\subsection{Hypotheses}

\begin{itemize}
    \item \textbf{H1:}  The decisions that we observe will follow Prospect Theory. In particular, we anticipate that participants will be risk-seeking for small probabilities ($RRP < 0$) and risk-averse for large probabilities ($RRP > 0$)~\cite{kahneman2013prospect}.
    
    \item \textbf{H2:} We hypothesize that visualization design will influence decision-making. Previous research has shown that people are more risk averse when presented icon arrays compared to text ~\cite{ruiz2013}. Therefore, we anticipate that $RRP_{icon} > RRP_{none}$.
    
\end{itemize}

\begin{figure}[b!]
    \centering
    \includegraphics[width=\linewidth]{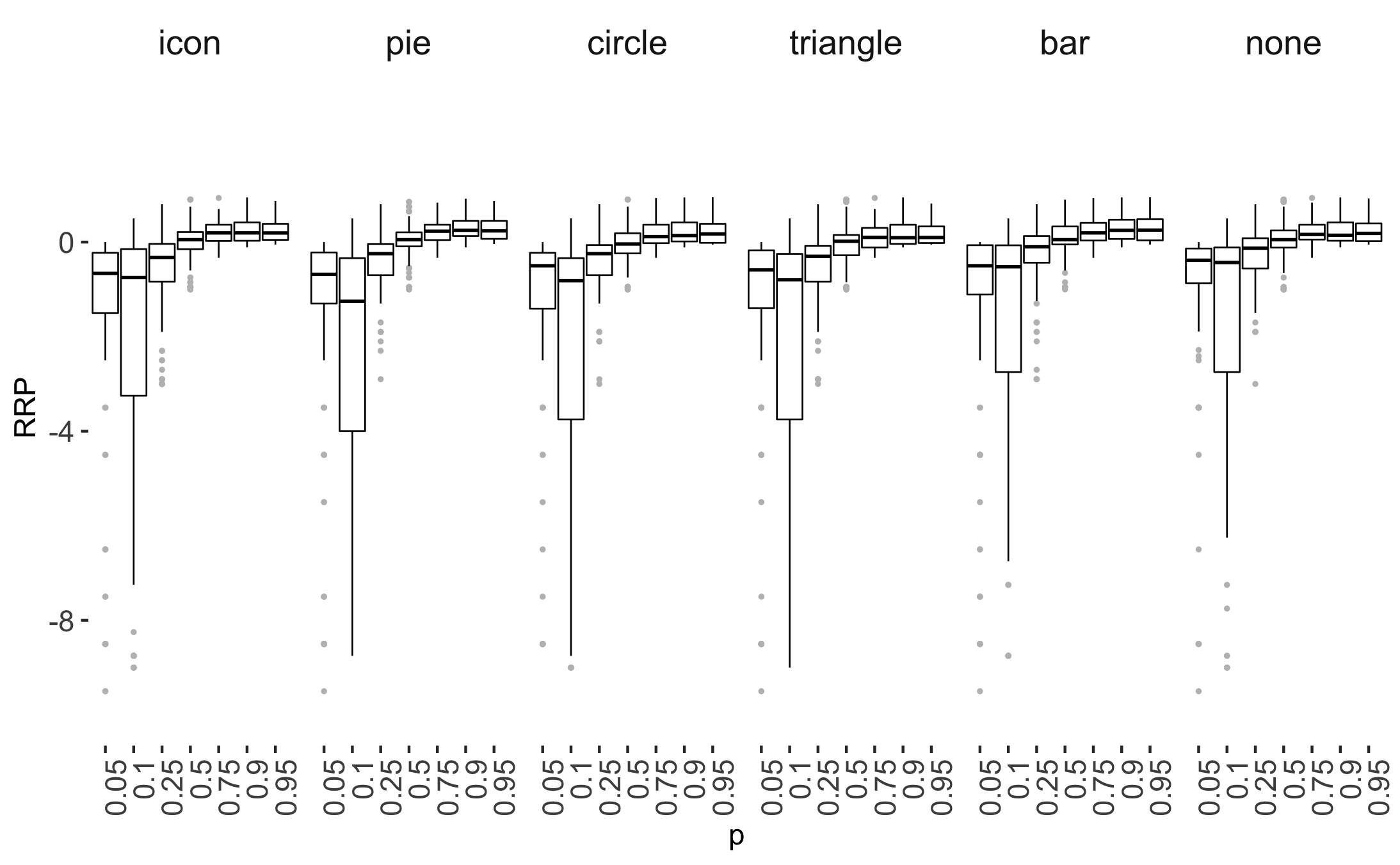}
    \caption{The RRP results showed that participants were risk seeking for small probabilities and risk averse for large probabilities.}
    \label{fig:box-rrp2}
\end{figure} 


\subsection{Results}


Our analyses utilized RRP to investigate the impact of visualization on the quality of lottery decisions. Figure~\ref{fig:box-rrp2} shows the distribution of RRP across different probability values and charts. We observed overall median \textsc{RRP} values of 0, 0.0155, 0.0273, -0.6625, -0.0500, and 0.0167 for \textit{icon}, \textit{pie}, \textit{bar}, \textit{triangle}, \textit{circle}, and \textit{none} respectively. The results show a similar behavioral pattern across all of the visualization conditions. Confirming H1 and replicating prior work~\cite{bruhin2010risk}, we found that participants were risk seeking for low probabilities (RRP~\textless~0) and risk averse for high probabilities (RRP~\textgreater~0).

We ran separate Kruskal-Wallis H non-parametric tests to determine whether the population medians for each of the probability values were dependent on the visualization types. The overall model revealed significant differences between the groups' \textit{RRP} distributions for each probability values, $\chi^2(4, N=8475) = 3408.1, p = 2.2e-16 $ and  $\eta p^2 = .05 $. Then, we ran separate Wilcoxon Mann-Whitney tests with Bonferroni-adjusted alphas  for visualization conditions across all probability values. We found that the risk-taking behaviors for participants in the \textit{circle} and \textit{triangle} groups deviated significantly from the \textit{none} group, \textit{suggesting that visualization can influence decisions}, thus confirming H2.

\subsubsection{Regression Model for RRP}

\begin{figure}[t!]
    \centering
    \includegraphics[width=\linewidth]{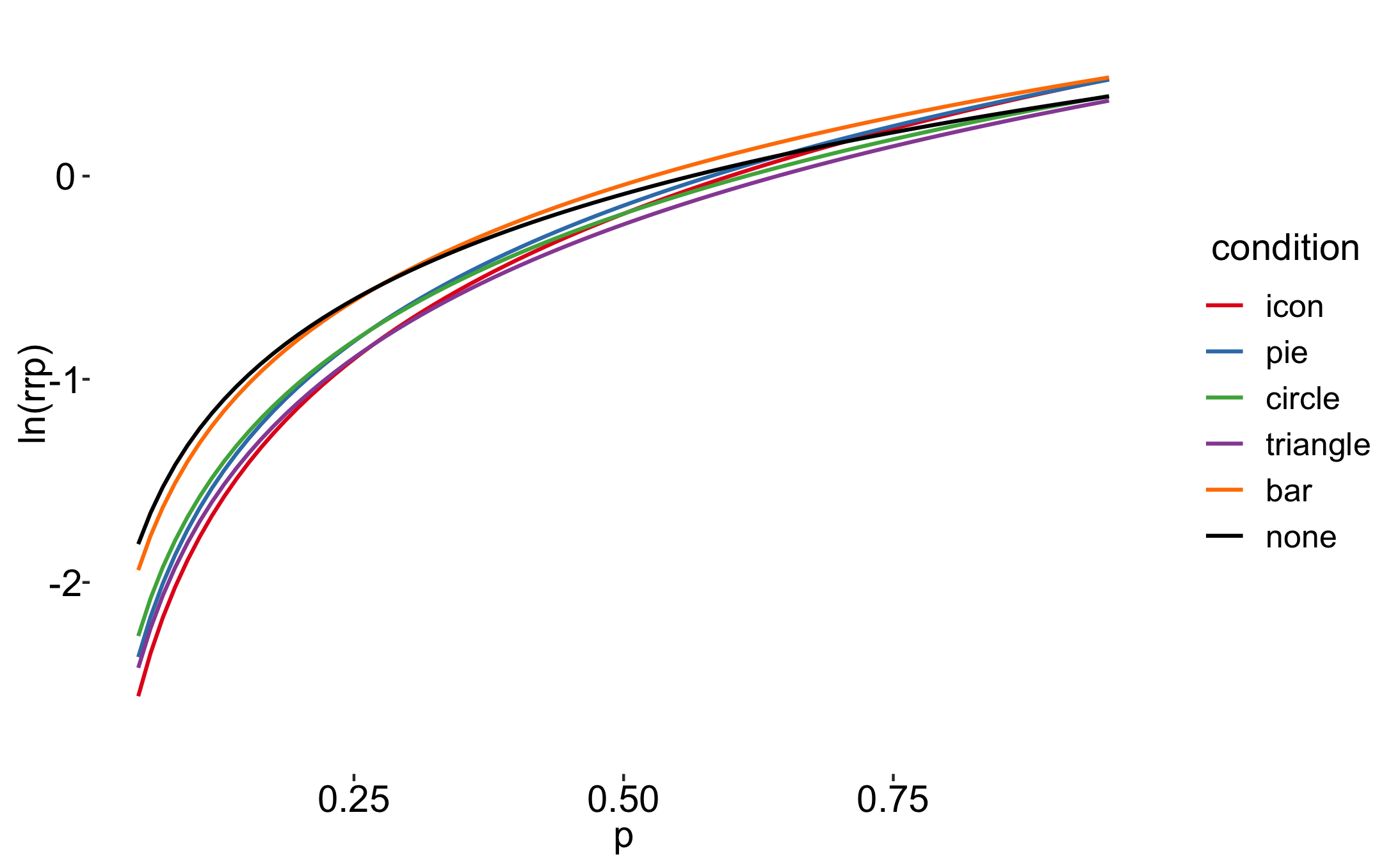}
    \caption{Logarithmic model for RRP across p. We observe that small probabilities lead to risk seeking behavior and large probabilities lead to risk averse behavior. Overall, the bar condition is the most similar to the text condition
    }
    \label{fig:logarithmic-rrp}
\end{figure}

We determined based on AIC and deviance that the logarithmic model was the best fit for every chart~\footnote{Our analysis also examined a simple linear regression. Refer to our repository \url{https://github.com/washuvis/letsgamble}.}. The logarithmic model takes the form:
\[
    y_v = \beta_{v0} + \beta_{v1}\ln(x),
\]
where $y$ represents the outcome variable and $x$ is the true probability. All models were significant with $p<001$, providing evidence that $p$ affects \textsc{RRP} (see Figure 4). The final regression models were:

\begin{itemize}[noitemsep]
    \item \textit{icon: $0.5296 + 1.0313\ln(p)$}
    \item \textit{pie: $0.5243 + 0.9653\ln(p)$}
    \item \textit{circle: $0.4411 + 0.9029\ln(p)$ }
    \item \textit{triangle: $0.1353 + 0.8828\ln(p)$}
    \item \textit{bar: $0.5281 + 0.8236\ln(p)$}
    \item \textit{none: $0.4307 + 0.7484\ln(p)$}
\end{itemize}
 
\subsection{Discussion}


Across all conditions, participants were risk-seeking with low-probability values and risk-averse with high probabilities. This replicates prior findings in  the economics domain \cite{bruhin2010risk} and confirms Prospect Theory \cite{kahneman2013prospect}. Moreover, we showed that risk behavior varies across different visualization designs.

Among the designs in this study, we found that \textit{icon} led to the least deviation from risk neutrality, with a median RRP of 0. However, our regression models suggest that participants in the \textit{bar} group exhibited behavior that was most similar to the \textit{none} group. These findings highlight the tension between the two possible interpretations of our results. On the one hand, according to economic theory, to maximize expected utility one should be risk neutral. Thus in the context of monetary decision-making, we show that \textit{icon} was most likely to elicit risk neutrality and is therefore the most effective design. On the other hand, the goal of the study is to understand the impact of the visualization on monetary risk behavior, thus implying a comparison to the control (\textit{none}) condition which represents expected behavior. We show that presentation can influence risk behavior, and demonstrate that the \textit{circle} and \textit{triangle} groups deviated significantly from the \textit{none} group. However, there is no consensus for what constitutes sound decision-making, and the preferred visualization may be context-dependent.

Our results only partially confirm the findings of prior work that indicated that icon arrays may lead to more risk-averse behavior~\cite{ruiz2013}. We found that participants in the \textit{icon} group were markedly more risk averse than the \textit{none} group when $p = .05$, but we observed no significant difference overall. This may be due to the contextual differences in the experiment designs and measures of decision making
Future work is needed to investigate contextual effect on risk taking behavior.

\subsubsection{Design Implications}

We began our investigations by drawing inspiration from designs used in the medical community.
In the context of medical risk, there is no
consensus on what constitutes a good medical decision, nor how to evaluate them~\cite{hamilton2017good}. In addition, considering the vast situational differences between an online user study and real-world medical decision-making, we hesitate to generalize our results. Still, the findings in this paper provide anecdotal evidence of the impact of the visualization on risk taking behavior. These results are particularly relevant given the increased popularity of visualization use by the general public to monitor the COVID-19 pandemic. Shared decision-making is also common practice, and patients make decisions based on risk factors, often using designs similar to the ones in this paper. 

Our experiment indicates that \textit{triangle} and \textit{circle} elicited risk-seeking behavior with the greatest deviation from risk neutrality. They were the only two conditions for which participants' behavior differed from the control group (\textit{none}). Our results provide immediate considerations for researchers and practitioners.

\subsubsection{Limitations and Future Work}

Although RRP provides a convenient measure for us to classify risk taking behavior, we hesitate to draw conclusions about the \textit{quality} of the decisions that we observed. The desirability of the risk-seeking or risk-averse behavior is context dependent. Moreover, participant bias needs to be taken into account. Still, risk neutrality is an objective measure of rational decision-making, and using RRP allows us to quantify and model departures from risk neutrality. 

One limitation of the study is that we only used gambling scenarios as a measure of the decision-making.
We selected gambling because it is a basic judgment task similar to the ones that people typically perform in everyday situations. However, we did not explicitly control for participant level differences and predisposition to gambling. Further investigations are needed to replicate this finding with differing pools of participants and different scenarios. 

Another limitation is the anchoring effect caused by certain probability values such as 0.25, 0.5, 0.75, and the potential bias imposed by using the donut chart for the instructions on visualizations such as \textit{pie}, that facilitated decoding. An extension of this study would be to conduct the experiment with intermediate values to minimize the anchoring effect. To support future investigations, all data are available at \url{https://github.com/washuvis/letsgamble}.

\section{Conclusion}
An analysis of gambling decisions found some consistent behavioral patterns across all visualization
conditions. We found that participants were risk-seeking with low-probabilities and risk averse with high probabilities. Furthermore, our results show a significant effect of visualization on risk attitudes. For instance, subjects in the circle and triangle groups exhibited greater risk-seeking behavior on average and all visualizations exhibited greater risk-seeking behavior compared to the text condition. 

 \acknowledgments{
 The authors wish to thank Joshua Landman for his insights on the statistical analyses. This material is based upon work supported by the National Science Foundation under grant number 1755734.}

\bibliographystyle{abbrv-doi}

\bibliography{template}
\end{document}


\begin{abstract}
    We address previous work, reviews to previous submissions and include additional analyses.
\end{abstract}
\maketitle

\section{1 Prior Work}
\subsection{1.1 Graphical Perception}

Our prior work (Icons are Best: Ranking Visualizations for Proportion Estimation. Zhengliang Liu, Melanie Bancilhon, Alvitta Ottley) in poster format (attached) investigates the effect of visualization on graphical perception. By prompting participants for numerical estimates of the proportions they saw in the visualizations, we compared the probability distortion across our five visualization designs. Overall, we found that the most accurate designs were icon\textgreater pie\textgreater bar\textgreater triangle\textgreater circle. Moreover, we found that participants tended to underestimate probabilities and overestimate large probabilities. Based on the above findings, we can infer that the differences in risk perception and decision-making across visualizations are closely \textbf{correlated} to the differences in graphical perception. This relationship is reinforced by a correlation score of \text{--}$0.327$ between RRP and BIAS (estimation \text{--} true value). 

For the \textbf{Graphical Perception Experiment}, we first conducted a pilot study for the graphical perception experiment with 300 participants (E1a). We conducted a subsequent similar experiment with the same participants, and added 106 new participants to the batch (E1b). For the prior work above, we used the data from E1b. 

\subsection{1.2 Graphical Perception + Decision-Making}

In this prior work, we were initially interested in investigating risk-behavior and decision-making. After the proportion estimate task, there was a second step in which participants were asked to make a lottery choice to gain a certain amount of money based on the probability displayed. We decided not to present the results from our second research question because we hypothesized that participants might have been biased by seeing their own numerical estimate. We developed another experiment (E2), where we assessed participants' decisions in a direct way, without prompting them for an estimate.

We combined our findings for both graphical perception and decision-making and presented our results in two other submissions. 

\subsubsection{Submission 1 Data}
In our first submission, we used the data from E1b and E2.

\subsubsection{Submission 2 Data}
In this second submission, we discarded the data for the 106 additional participants in E2b. The reason for our decision was to have equal sample sizes for both Experiment I and II. We compared the pilot study and the subsequent study and found that the results were consistent

\section{2 Additional Analyses}

\begin{figure} 
    \centering
    \includegraphics[width=\textwidth]{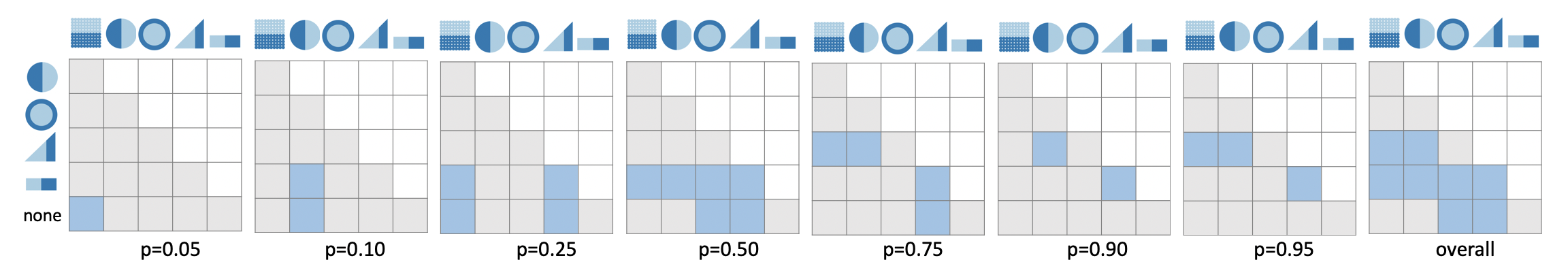}
    \caption{Pairwise comparisons for RRP. Blue squares show significant difference and grey squares show no significant difference. Our overall model shows that participants in the \textit{circle} and \textit{triangle} groups deviated significantly from the \textit{none} group, suggesting that visualization can influence risk behavior. }
    \label{fig:pairwise-rrp}
\end{figure}

\subsection{2.1 Pairwise Comparisons}

Figure 1 shows the pairwise comparisons using Kruskal Wallis tests.

\subsection{2.1 Logarithmic Regression}
Table 1 shows the detailed results for the rsquare, AIC, skewness, kurtosis and deviance values for the RRP logarithmic regression model for each design.

 \begin{table}
\caption{The rsquare, AIC, skewness, kurtosis and deviance values for the RRP logarithmic regression model for each design.  These values were compared to a simple linear regression and showed a better goodness-of-fit based on AIC and deviance values.}
\label{tab:logarithm_model}
\resizebox{\linewidth}{!}{
\begin{tabular}[.4\textwidth]{|cccccc|}
\hline
$condition$  & $rsquare$ & $AIC$ &  $skewness$   & $kurtosis$  & $deviance$ \\ 
\hline

icon & 0.21  & 6996.28 & -4.52 & 27.20 & 7396.81\\
pie & 0.25 & 4869.07 & -3.79 & 21.53 & 3799.68  \\
circle & 0.23  & 3960.95 & -4.04 & 24.11 & 3074.05\\
triangle  & 0.20 & 5476.71 & -4.35 & 26.55 & 5117.81\\
bar & 0.20  & 4841.80  & -5.02 & 39.10 & 3716.02	 \\
none & 0.18  & 4619.10  & -5.37 & 39.27 & 3488.37	 \\

\hline
\end{tabular}
}
\end{table}